\documentclass[prd, twocolumn, nofootinbib]{revtex4}
\usepackage{graphicx}

\newcommand{\beq}{\begin{equation}}
\newcommand{\eeq}{\end{equation}}
\newcommand{\beqa}{\begin{eqnarray}}
\newcommand{\eeqa}{\end{eqnarray}}
\newcommand{\om}{\Omega_m}

\def\fun#1#2{\lower3.6pt\vbox{\baselineskip0pt\lineskip.9pt
  \ialign{$\mathsurround=0pt#1\hfil##\hfil$\crcr#2\crcr\sim\crcr}}}

\begin{document} 

\title{Cosmological Model Selection: Statistics and Physics}
\author{Eric V.\ Linder} 
\affiliation{Berkeley Lab, University of California, Berkeley, CA 94720, USA} 
\email{evlinder@lbl.gov}
\author{Ramon Miquel}
\affiliation{Instituci\'o Catalana de Recerca i Estudis Avan\c{c}ats,
 Institut de F\'{\i}sica d'Altes Energies\\
 Campus UAB, E-08193 Bellaterra (Barcelona), Spain}
\email{rmiquel@ifae.es}
\date{\today} 

\begin{abstract} 
Interpretation of cosmological data to determine the number and values 
of parameters describing the universe must not rely solely on statistics 
but involve physical insight.  When statistical techniques such as ``model 
selection'' or ``integrated survey optimization'' blindly apply Occam's 
Razor, this can lead to painful results.
Sensitivity to prior probabilities and to the number of models compared 
can lead to ``prior selection'' rather than robust model selection. 
A concrete example demonstrates that Information Criteria can in fact 
misinform over a large region of parameter space. 
\end{abstract} 

\maketitle 

\section{Introduction} \label{sec:intro} 

Recently several papers (e.g.~\cite{liddle0701,liddle0610,trotta}, 
and references therein) have claimed that fitting cosmological 
parameters to data may be inadequate, if not misguided, as a means of 
determining cosmology, in particular the nature of dark energy accelerating 
the expansion of the universe.  They advocate interpreting the evidence 
for or against a model in a manner intimately tied to the parameter 
space volume: models with fewer parameters get boosted in preference 
while models with more parameters -- and possibly better fits -- should 
be penalized.  This is all in accord with statistical theory.  In this 
approach one should not fit parameters but rather select models; we 
will refer to the various related techniques along these lines generically 
as ``model selection''. 

However, this point of view can be taken too far, and has as well 
some essential assumptions that prevent model selection from being 
either a panacea or a substitute for parameter fitting (which admittedly 
can have pitfalls of its own).  
In general, statisticians want to model data in an efficient way. 
That is, they want to go from many data points to just a few parameters 
in a model that captures the essence of the data accurately and efficiently. 
Then, one can use that model to predict future outcomes, say 
(clinical trials, stock market, climate change, etc.). 
However, physicists do not regard the models as just useful
summaries of the data, but as fundamental descriptions of the data based
on physical principles. The parameters in our models have (or should 
have) deep physical
meanings, and are not just concise ways of representing the
data. Model efficiency takes a back seat to physical fidelity.

We discuss drawbacks of the model selection paradigm 
in \S\ref{sec:prior}, and give several historical examples in 
\S\ref{sec:history} where model selection would have led physicists 
astray, especially as a predictive mechanism.  In \S\ref{sec:example} 
we present a complete numerical example, using simulated cosmological 
data, that shows how easily model selection can lead us to the wrong 
physics.  Inadequate data, or extrapolation to future, 
better data, is particularly problematic within a model selection 
scenario since that rewards mathematical simplicity, which blurry 
data are more likely to be consistent with.  We emphasize in 
\S\ref{sec:future} that one can too easily misuse this paradigm in 
attempting survey design, judging data before it is taken. 

This article does not oppose model selection as an approach, it just provides 
necessary highlights of areas of caution such as sensitivity to the outer 
ranges -- ``corners'' -- of priors and unphysical dependence on number of 
models being considered.  Through a series of concrete examples we illustrate 
why physics, not just statistics, must retain a central role in 
interpreting data. 

\section{Bayesian Approach} \label{sec:prior} 

An essential ingredient of the parameter fitting paradigm is 
that the model described by the parameter set is a reasonable choice. 
Model selectors ask what if the model is faulty to begin with, and 
seek to avoid, or at least dilute, this step by considering sets of models, 
containing parameters, rather than sets of parameters.  As far as this 
goes, it is certainly reasonable -- parameter fitters must take care 
in choosing parameter sets, with respect to validity and lack of bias.  
However, in parameter fitting to 
data one already has built in tests to check that the model is 
reasonable: 1) ideally one starts with a physically motivated model, 
2) if the goodness of fit (e.g.\ chi-squared per degree of 
freedom) is poor then the model receives careful scrutinization or 
is discarded, and 3) if subsets of data disagree then this may be a 
sign that the parameter space should be expanded or abandoned.  So 
parameter fitting can check itself. 

Model selection seeks to consider a space of models and find the 
Bayesian evidence for each.  Here the key ingredient is the assigning 
of prior probabilities to parameter states within each model.  That is, 
in a not-insignificant sense one must guess what the data are going to 
say before you have them. 
The final answer from model selection depends on the prior assumptions, 
assumptions often with relatively little physics guidance.  Whether the 
prior is uniform in a parameter itself or in a function of the parameter 
has substantial impact on the results (for one telling example concerning 
our very existence, see \cite{starktrotta}).  The range of the prior is 
another crucial element.  Finally, while in parameter fitting one must ensure 
that the particular parameter values of the fit are physical, in model 
selection one must extend the probability analysis throughout the entire 
range of parameter values, some portions of which may be pathological 
without this being apparent (indeed the ``corners'' -- the extreme ranges 
of the prior -- constitute 99.8\% of the volume in a 10 dimensional space 
\cite{loredo}).  Only after these assumptions can the model 
probability be computed.  

The final quantity for selecting models is their evidence value, the 
mean likelihood averaged over the prior.  Since probability must sum 
to unity, large parameter spaces, or spaces where the prior extends 
widely, are strongly penalized.  Doesn't this make sense, though? 
Shouldn't we apply Occam's Razor to shave the models with more ``hair''? 
We agree that the consistency of data with a model with few parameters 
can be a useful guide, but there are several points that caution 
against overly strict application of model selection: shaving blindly 
with Occam's Razor is a dangerous activity. 

For one thing, there is only one universe accessible to us, so 
prior probability has a very limited meaning.  If only one point in 
the model parameter space fits the data, nevertheless our universe 
may be described by that model.  Indeed, even the number of 
models being compared influences the probabilities, as the Bayesian evidence 
is calculated under the constraint that the total probability must sum 
to unity. 

For another, any model that somewhere fits the data must be allowed 
to live or else we sacrifice the good with the bad.  
If there is one ``fit'' fruit on a tree with bad apples, using model 
selection criteria everything would be thrown away without exception. 

The role of the prior for model selection is crucial, yet the method 
of selection of the prior is completely undefined.  Consider the 
equation of state phase space $w$-$w'$, where $w$ is the equation of 
state ratio of the dark energy pressure to energy density, and 
$w'=dw/d\ln a$.  Should we 
choose a prior distribution uniform in $w$ and $w'$, or perhaps 
uniform in $w$ and $\dot w=Hw'$ -- the Bayesian evidence will be 
different in the two cases and models may be ruled out in one case 
but not the other.  Or should we choose a uniform prior in the field 
values $\phi$ and $\dot\phi$? Perhaps $\dot\phi^2$ and $V$, where $V$ 
is the potential energy and $\dot\phi^2/2$ the kinetic energy?  Perhaps 
$V$ and $V'$, where $V'=dV/d\phi$?  What model selection tells us 
is an acceptable cosmology differs in each case.  The foundation, 
the priors, are essentially undefined by physics but put in by the 
model selectors by hand.  Indeed it is tempting to call this approach 
``prior selection'' rather than model selection.  Many of these 
issues are indeed recognized in the statistics community (see, e.g., 
\cite{jefferys}), if not with full emphasis on physical fidelity. 

Mathematical simplicity -- the essential weighting for model selection -- 
does not necessarily accord with physical simplicity.  We do not always 
recognize simplicity when we see it.  Consider the model spanned by the 
coefficients of a polynomial function $f(x)=a_0+a_1 x+a_2 x^2+\dots$ 
Model selection will deal harshly with it, despite that the apparent 
complexity -- in some cases an infinite number of parameters -- 
sometimes represents simplicity, such as $f(x)=e^x$, with zero parameters.  
Conversely, models 
which appear simple can hide deeper complexity, e.g.\ where a Gaussian 
is actually approximating the convolution of multiple functions. 

Braneworld models in a spatially flat universe are extremely simple, 
involving one parameter, the crossover distance.  However if we do 
model selection in terms of the distance-redshift relation $d(z)$, 
or the equation of state ratio $w(z)$, it appears the braneworld 
models have many more parameters than the cosmological constant 
model.  Model selection will penalize strongly the braneworld 
model for the {\it apparent\/} number of parameters, despite the 
underlying physics being quite simple.  As another example, a 
constant equation of state model would appear to have many parameters 
if discussed in terms of redshift-binned densities $\rho(z_i)$ or 
principal components of the density (also see \cite{sahnistaro}).  
Thus the concept of simplicity is actually quite complicated. 
We discuss this further in \S\ref{sec:history}, giving historical examples. 

Parameter fitting allows us to compare cosmologies equally in the 
space describing cosmological distances, or densities, or equation 
of state, or 
somewhat ``non-parametric'' forms like principal components.  Model 
selection imposes an unphysical overlayer of expectation on such 
comparisons, based on a perceived simplicity. 

Parameter fitting does not require prior knowledge of the probability 
weighting within the parameter space; it allows the physics to, 
literally, impose ``survival of the fittest''.  The most useful 
parameterizations are moreover motivated physically.  Consider the 
equation of state phase space $w$-$w'$.  While model selection would 
weight this as a whole, as two parameters, physics calls out certain 
subregions of this space as being especially physically relevant 
\cite{caldlin}.  In particular, thawing dark energy cosmologies,  
where the dark energy was locked by Hubble expansion drag in a 
cosmological constant like state in the early universe and more 
recently released to evolve, follow one-dimensional trajectories in 
a narrow region, or inverse power law models actually possess only a single 
equation of state parameter, not two. 
Model selection deweights these solutions because -- although they may 
be perfectly good fits -- they reside in a larger 
parameter space.  Conversely, model selection favors constant equation 
of state models, with a single parameter, despite their being 
physically unstable, requiring a precise fine tuning of the ratio 
between kinetic and potential energies.

\section{Reality Check} \label{sec:history} 

Prior to 1998, supernova distance data were consistent with a flat, 
matter dominated universe.  Model selection would have claimed 
strong evidence for the zero parameter SCDM ($\om=1$) model, 
disfavored the general matter dominated universe OCDM ($\om\in[0,2]$, 
say), and dismissed a cosmological constant universe $\Lambda$CDM ($\om$ + 
$\Lambda$).  Simplicity does not automatically bring truth. 

Feynman famously pointed up the pitfall in calling on simplicity as 
a guidepost in physics by rewriting all of physics as $\vec U=0$, 
where each element of $\vec U$ contained the hidden structure, e.g.\ 
Newton's second law as $U_1=F-ma$. 

Before 1992 the Bayesian evidence was overwhelming against structure 
in the cosmic microwave background, i.e.\ comparing the model 
parametrized by the temperature power spectrum multipoles $C_\ell=0$ 
vs.\ the much larger parameter space spanned by $C_2$, $C_3$, etc. 
But no one would have taken such an argument seriously.  (Of course 
using individual $C_\ell$ is a statistical approach not a physical one, 
but that is part of the point about needing physical input.) 

The galaxy two-point spatial correlation function for many years was 
thought to be a power law, with only two parameters, yet now we know 
there is little physics in that model; rather the apparent power 
law is nearly a coincidence from evolution in the true matter clustering and 
in the bias between mass and light.  

The mass power spectrum looks like a smooth curve yet we appreciate 
the deeper physics only by going to an apparently more complicated halo model 
involving the addition of multiple terms. 

In a particle physics context, the Fermi theory of weak interactions 
managed to describe all low-energy phenomenology with one single parameter, 
the Fermi constant $G_F$. However, the underlying renormalizable physical 
theory, the Glashow-Weinberg-Salam Standard Model of electroweak 
interactions, has many more free parameters, and has also a realm of 
application much wider than that of the Fermi theory, which came to be 
regarded as just a low-energy effective theory.

These historical examples are clearest for information criteria 
approximations to the full Bayesian evidence, but such simplicity or 
``predictiveness'' arguments apply to the evidence as well.  While the 
full evidence factorizes out unconstrained parameters, these 
parameters may not be recognized, e.g.\ when an equation of state parameter 
is viewed in density or distance (see \S\ref{sec:prior}), 
and in any case the mere presence of more 
models affects the evidence. 

Purely statistical ideas of simplicity are insufficient in physics. 
To a large extent, beauty follows truth, not truth beauty.  What we 
consider simple is what we understand.  A mess of spectral lines from 
a hydrogen atom is ugly until we understand the physics behind it, then 
we realize it is beautiful and simple.  The simple cosmological 
constant hides, within present physics knowledge, all sorts of ugly 
fine tunings, while a complicated equation of state $w(z)$ may turn into 
something as beautiful as the hydrogen atom.  Model selection forces us 
to give undue weight to our guess as to which is an ugly duckling and 
which will be a beautiful swan.

\section{A Worked-Out Example} \label{sec:example}

As a specific example of the overemphasis that model selection places 
on models with fewer parameters, despite the physics, we analyze 
constraints on dark energy from future data.  For such data we take 
a distance-redshift survey over redshifts $z=0.1-1.7$, similar to the 
supernova half of the proposed SNAP satellite mission~\cite{SNAP}, 
supplemented by 300 local Hubble flow ($z=0.03-0.08$) supernovae as 
will be provided from the Nearby Supernova 
Factory (SNF) survey~\cite{SNF}, and by measurement of the distance to 
the surface of CMB last scattering provided by the ESA-NASA
mission Planck~\cite{Planck}. 

The supernovae (SNe) numbers and redshift distribution are as given 
in~\cite{KLMM}.  We include a linearly rising 
systematic error floor in each $\Delta z = 0.1$ redshift bin 
of $\sigma_{\mathrm{syst}} = 0.02z/1.7$, to be added in quadrature in 
each redshift bin to the measurement error given 
by $\sigma_{\mathrm{int}} / \sqrt{N_{\mathrm{SN}}}$, with 
$\sigma_{\mathrm{int}} = 0.12$ and $N_{\mathrm{SN}}$ being the number 
of Type Ia SNe in that bin.  We take a relative precision of
0.7\% in the Planck distance measurement.

A standard cosmological fit analysis of actual SNAP + SNF + Planck 
data will produce a central value in the ($w_0$,$w_a$) plane (recall 
that $w_a = -2w'(z=1)$) and a contour around it encompassing some 
chosen confidence level (CL), typically 68, 90, or 95\%. The point 
corresponding to a cosmological constant (-1,0) may or may not lie 
inside the contour. If it does not, we may say that we 
exclude $\Lambda$CDM at that CL. 

\begin{figure}[!htb]
\includegraphics[width=3.5in]{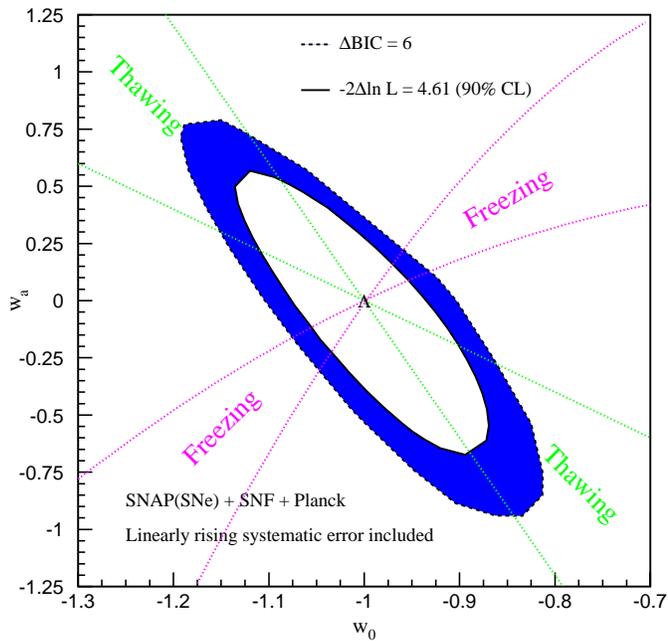}
\caption[Confidence level contours]{
Confidence-level contours for a SNAP(SNe) + SNF + Planck combined data 
sample. The cosmological models defined by ($w_0$,$w_a$) lying outside 
the inner contour would be favored over $\Lambda$CDM at more than 
90\% CL by a standard fitting analysis. The models lying inside the 
outer contour would be strongly rejected by a BIC analysis. 
The shaded area in between the contours corresponds to the loci of 
models that, if realized in nature, 
would be preferred at over 90\% CL by fitting, but strongly, 
and wrongly, rejected by BIC. 
\label{fig:contours}
}
\end{figure}

More precisely, in a frequentist analysis in which one aims to prove 
or disprove $\Lambda$CDM, one would simulate the expected SNAP + SNF + 
Planck data sample assuming a $\Lambda$CDM universe, and, analyzing this 
synthetic data sample as if it were the real data, one would draw 
a, say, 90\% CL contour around the (-1,0) point, much like the 
one depicted as the inner contour in Fig.~\ref{fig:contours}. 
(Note no assumption of Gaussianity in likelihoods has been made.) 
That contour tells us that, if $\Lambda$CDM is true, we expect to get 
a central value inside that contour in 90\% of the observations like 
SNAP + SNF + Planck that we may perform.  Therefore, if our one real 
SNAP + SNF + Planck observation delivers a central value outside that 
contour, irrespectively of its associated error ellipse, we will be able 
to say that we have excluded $\Lambda$CDM at greater than 90\% CL.  
For all true models defined by ($w_0$,$w_a$) values outside that contour, 
we expect (in the statistical sense) the measured values to lie, indeed, 
outside, and so we expect to be able to select them over $\Lambda$CDM at 
greater than 90\% CL. 

With the information criteria approach used by model selection advocates, 
one penalizes models with more parameters.  Various information criteria 
exist (see \cite{liddle0701} for an overview); here we concentrate on 
the Bayesian Information Criterion (BIC), which is closely related to 
the Bayesian evidence but frequently used because it is simpler to compute.  
The outer contour in Fig.~\ref{fig:contours} shows the 
formal BIC contour corresponding to $\Delta{\rm BIC}\equiv
{\rm BIC}(w_0,w_a) - {\rm BIC}(\Lambda{\rm CDM}) = 6$. 
Values of $\Delta{\rm BIC}\ge 6$ indicate strong evidence against the 
model being probed. That is, for all models defined by ($w_0$,$w_a$) 
inside the outer contour of Fig.~\ref{fig:contours}, BIC would give 
strong evidence against them and strong evidence supporting, instead, 
$\Lambda$CDM.  (In contrast, the 
iso-contour in likelihood L with  
$-2\Delta \ln {\mathrm L} = 9.48$ (not shown), which, with the penalty from 
$(w_0,w_a)$ models having two additional parameters beyond $\Lambda$CDM, 
roughly corresponds in our problem to $\Delta$BIC = $6$, yields a 
99.13\% CL {\it against\/} $\Lambda$CDM in the fitting analysis.)  

For this combined data sample, one can say that a Bayesian Information 
Criterion analysis misinforms us about, i.e.\ spuriously rules out, 
all models inside the dashed contour.  
In order for BIC to give the same hint against $\Lambda$CDM that the 
standard fitting analysis gives at 90\% CL, one has to go in this case 
to a contour with $\Delta$BIC = $-2$ (or roughly 
$-2\Delta \ln {\mathrm L} = 17.48$), 
which would extend to the margins of the figure. 
Note $-2\Delta \ln {\mathrm L} = 17.48$ corresponds to 99.98\% CL in the 
fitting analysis. That is, only when the fitting analysis gives $\Lambda$ 
less than a $2\times10^{-4}$ chance of being correct, does model 
selection start to give a weak hint of preference for the $(w_0,w_a)$ 
model, turning away from the incorrect model 
$\Lambda$CDM. 

For all models whose ($w_0$,$w_a$) lies between the two contours in 
Fig.~\ref{fig:contours} the standard fitting analysis would guide us 
away from $\Lambda$CDM, while the BIC analysis would strongly, and wrongly, 
discard the true -- by construction -- model.  
Such are the dangers of model selection based on simplicity arguments. 
Indeed the area of the BIC ``misinformation'' 
region is larger than the entire region where $\Lambda$ is viable according 
to standard fitting. Fig.~\ref{fig:contours} also shows the physically 
motivated freezing and thawing regions \cite{caldlin} where models 
should generically lie.  We see that BIC is particularly slanted against 
the half of the physical space that is thawing models.

\section{Prediction and Survey Design} \label{sec:future}

As shown from the historical examples in \S\ref{sec:history}, model 
selection is unsuitable for predicting constraints from future data 
and surveys.  Physics input, not statistical probability, is required 
for assessing the value of future surveys.  

Claims for survey 
optimization that average likelihoods over prior probabilities are 
basically betting on absence of any physical structure.  A similar 
assumption was addressed in \cite{linbias} where it was shown that 
a figure of merit based solely on the area of the confidence region 
was robust only on abdication of physics input.  Such a ``blank map'' 
approach was there called Snarkian statistics, after Lewis Carroll's poem, 
``The Hunting of the Snark''.  Using model selection to design 
surveys or plan for the future based on present consistency with 
the cosmological constant is similarly like saying the snark hunters 
are well served by a ``map representing the sea without the least vestige 
of land'' because the sea so far has been featureless: 
it may be simple and favored in model selection, but is 
not to be trusted for navigating if reefs and islands -- structure -- 
may exist. 

Standard parameter fitting methods should and can test if the model framework 
is reasonable or not, as mentioned in \S\ref{sec:prior}. 
In particular, the cosmological $w_0$-$w_a$ parameterization 
was specifically invented from the physical motivation of scalar field 
behavior \cite{linprl} and has been carefully shown to be largely safe 
from bias \cite{linbias}.  Goodness of fit alerts us to problems in the 
parameterization, informs us if we should enlarge the parameter 
space (e.g.\ by comparison of the results of data subsets or 
multiple probes -- a very important aspect of learning about dark 
energy physics), and allows comparison of disjoint models that do not share 
a parameter space.  We do not need to know a priori the probabilities 
at every point in parameter space; the data lead us.

\section{Conclusions} \label{sec:concl} 

When coupled with strong physical inputs, tests of robustness 
against priors, and precise data, model selection techniques can 
be a useful companion to the standard parameter fitting approach. 
They are not, however, a panacea or a way to draw conclusions 
stronger than actual data can support.  We cannot get answers about 
the physics before we get the data.  The sensitivity to prior guesses 
about the probability of the correct physics runs the risk of turning 
this approach into ``prior selection'' rather than model selection. 

Overenthusiastic application of model selection has led to some 
claims about the probability of future experiments failing to see 
characteristics such as dynamics that current data cannot access.  
Such statistical pronouncements from the shaky foundation of priors 
are reminiscent 
of the Dao De Qing, which lessons us that ``Without going outside, you 
may know the whole world, without looking through the window, you may 
see the ways of heaven.  The farther you go, the less you know.''  
Though a larger parameter space -- representing ``farther'' 
physics -- has many points that 
fail to fit, only one point needs to be correct, and that large 
parameter space may arise from a simple, beautiful theory once we 
understand it. 
As the eminent statistician C.R.~Rao puts it, ``The [information] 
criteria do not take into account the purpose for which the model 
is estimated and the loss incurred in using the estimated distribution 
to predict future events'' \cite{rao}. 

Parameters should have physical meanings, and statisticians' model 
efficiency then takes a back seat to physical fidelity.  Through a 
concrete example we showed how a statistical Information Criterion 
can decisively rule out a physical model -- incorrectly.  Furthermore, 
the misinformation area of model selection exhibited in 
Fig.~\ref{fig:contours} grows even larger with more parameters. 
Recalling our Standard Model analogy of \S\ref{sec:history}, we 
should have no expectation that the physics of 75\% of the universe's 
energy density can be explained by a single number.

\section*{Acknowledgments}

We thank Don Groom, and Andrew Liddle for lively discussions.  This work 
has been supported in part by the 
Director, Office of Science, Department of Energy under grant 
DE-AC02-05CH11231.

\end{document}